# Impurity-induced magnetic order in the mixture of two spin gap systems $(CH_3)_2CHNH_3CuCl_3$ and $(CH_3)_2CHNH_3CuBr_3$

Keishi Kanada [a], Takehiro Saito [a], Akira Oosawa [a], Takayuki Goto [a]
Takao Suzuki [b,a] and Hirotaka Manaka [c]

[a] *Department of Physics, Faculty of Science and Technology, Sophia University 7-1 Kioicho, Chiyodaku Tokyo 102-8554, Japan*
[b] *Advanced Meson Science Laboratory, RIKEN 2-1 Hirosawa, Wako, Saitama 351-0198, Japan*
[a] *Graduate School of Science and Engineering, Kagoshima University Korimoto, Kagoshima 890-0065, Japan*

**Abstract**

The ground state of the solid solution of the two spin gap systems $(CH_3)_2CHNH_3CuCl_3$ and $(CH_3)_2CHNH_3CuBr_3$ has been investigated by $^1$H-NMR. The existence of a magnetic ordering in the sample with the Cl-content $x$=0.85 was clearly demonstrated by a drastic splitting in a resonance line at low temperatures below $T_N$=13.5K. The observed NMR spectra in the ordered state was qualitatively consistent with the simple antiferromagnetic state.

## 1. Introduction

The title compounds of $(CH_3)_2CHNH_3CuCl_3$ and $(CH_3)_2CHNH_3CuBr_3$ abbreviated as IPA-$CuCl_3$ and IPA-$CuBr_3$ are 1-D spin gap systems with an isomorphic structure[1,2]. The mechanism and the magnitude of the gap are quite different in these two systems. In the Cl-system, which has recently been revealed to have a ladder-like magnetic structure by the inelastic neutron experiment [3], the neighboring two $S$=1/2 spins on the rung interact ferromagnetically to form effective $S$=1 spins at low temperatures. The interaction among these integer spins is antiferromagnetic, so that the ground state is gapped one as Haldane conjectured [4]. In Br-system, on the other hand, the interaction between the neighboring two $S$=1/2 spins is antiferromagnetic, so that they form a singlet dimer state at low temperatures. The spin excitation gaps of these two systems have been reported to be about 14 K and 98 K by neutron and magnetic susceptibility experiments [1,2,3]. The purpose of this study is to investigate the ground state of the solid solution of the two spin gap systems.

So far, Manaka *et al.* have been working intensively on the macroscopic measurements in IPA-$Cu(Cl_xBr_{1-x})_3$ to report that the magnetic order takes place within the limited region $0.44<x<0.87$, and that the transition temperature shows a jump from zero to finite values at the two boundaries [5,6]. Nakamura has investigated numerically on the spin system with a bond disorder to report that an antiferromagnetic ground state will be realized by a quantum effect [7].

*Email address:* gotoo-t@sophia.ac.jp (Takayuki Goto)

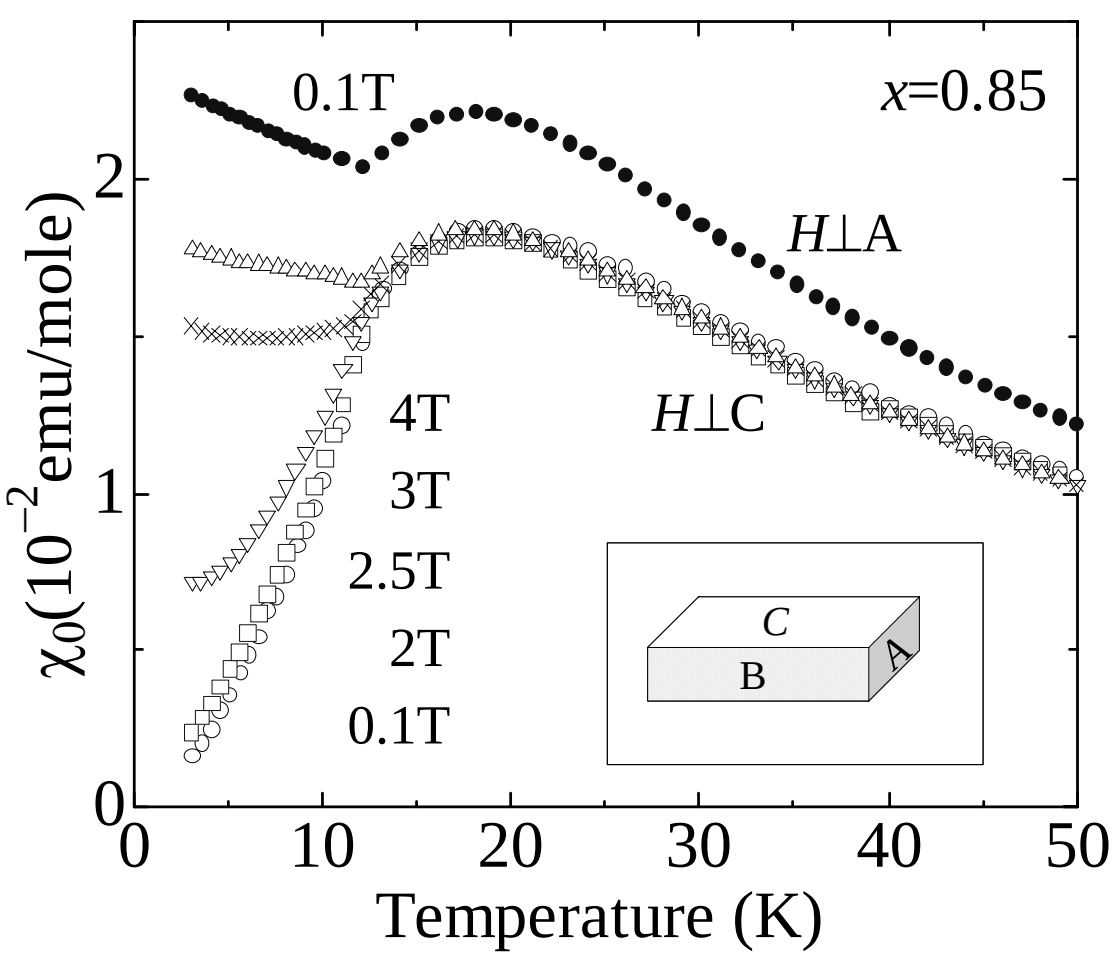

Fig. 1 The temperature dependence of the uniform susceptibility $\chi_0$ of the $x$=0.85 sample under fields between 0.1T to 4T. The inset shows the rectangular shape of a crystal, with a denotation of the three surfaces [1].

However, microscopic investigation on the ground state has been left untouched except for our preliminary result by μ-SR [8]. In this paper, we demonstrate the existence of a magnetic phase transition in IPA-Cu($Cl_xBr_{1-x}$)$_3$ ($x$=0.85) by $^1$H-NMR spectra which are qualitatively consistent with the simple antiferromagnetic spin structure with $\vec{q} = 0$.

## 2. Experimental

Single crystals of IPA-Cu($Cl_xBr_{1-x}$)$_3$ ($x$=1 and 0.85) were grown by an evaporation method from isopropanol solution of isopropylamine hydrochloride, copper(II) chloride dihydrate, isopropylamine hydrobromide, copper (II) bromide, contained in a bowl, which was maintained to be 30(±0.1)ºC and in a atmosphere of flowing nitrogen gas during an entire period of crystal grow, typically a month or two. A typical size of obtained crystals was around 3×5×10mm with a rectangular shape as was reported [1,2,6].

The content of Cl was determined by ICP method on three tiny fragments chipped off from different points of crystals. Macroscopic quantities of obtained crystals, the specific heat and the susceptibility were measured by PPMS and MPMS manufactured by Quantum Design Co. Ltd. $^1$H-NMR measurements were performed by using a 4K cryogen-free refrigerator set in a 6.5T cryogen-free-superconducting magnet. Spectra were measured by recording a spin-echo amplitude simultaneously with ramping the magnetic field. In measurements under a field of different orientations, the sample was rotated by a geared rotator.

## 3. Results and Discussion

Figure 1 shows the temperature dependence of the uniform susceptibility $\chi_0$ of $x$=0.85 sample under fields between 0.1T and 4T. At low field 0.1T, $\chi_0$ shows a kink around 12K, and below that temperature, it decreases steeply in the case of $H$⊥C-plane, while slightly increases in the case of $H$⊥A-plane [1]. The specific heat also showed a small but distinct cusp at the same temperature. These results indicating the existence of a magnetic long range order and the direction of the easy axis being perpendicular to C-plane are consistent with those reported by Manaka [5]. With increasing field, the behavior of $\chi_0$($H$⊥C) does not change until 2T, and at still higher fields the reduction below the kink temperature is weakened rapidly.

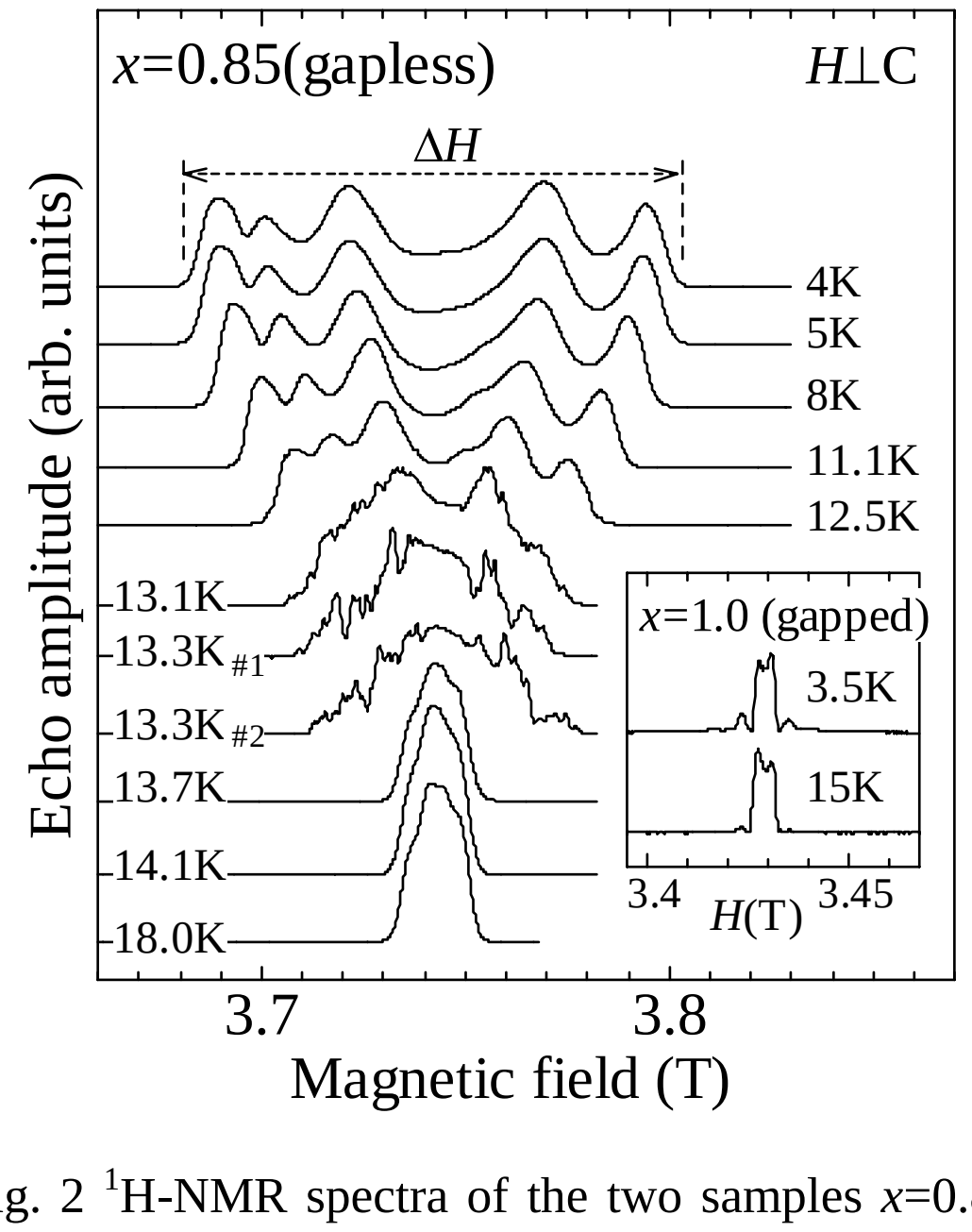

Fig. 2 $^1$H-NMR spectra of the two samples $x$=0.85 and 1 under the field at around 3T. The two spectra #1 and #2 at 13.3K are obtained by independent scans under the identical conditions.

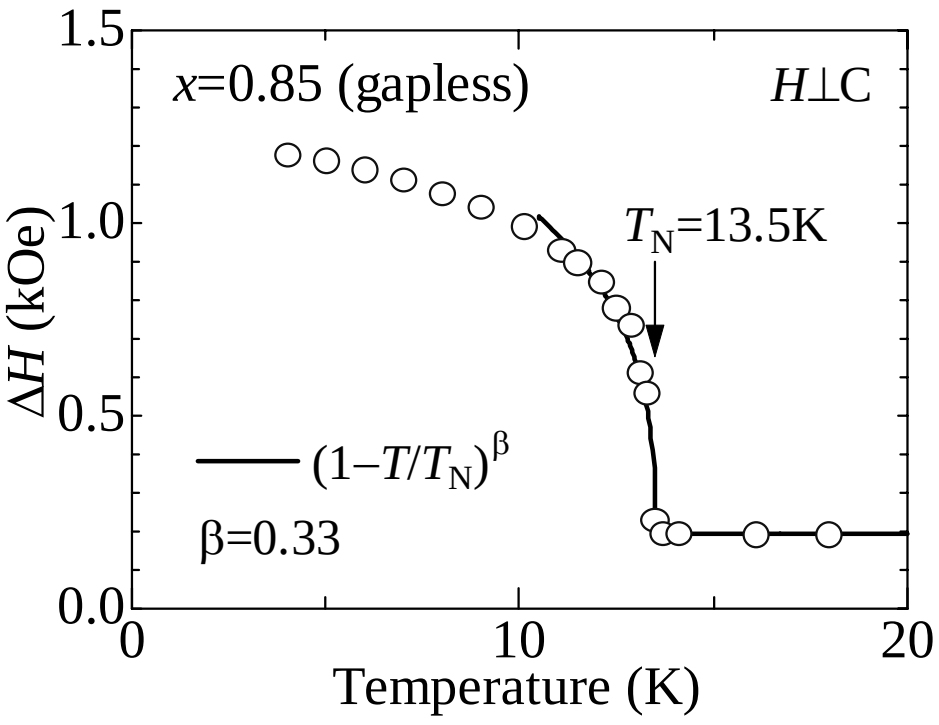

Fig. 3 Temperature dependence of $\Delta H$, definition of which is shown in Fig. 2. The value of $\Delta H$ above $T_N$ corresponds to the peak width in the paramagnetic state $\Delta H_0$=0.20 kOe.

This means that the spin flip transition takes place at a field slightly above 2T as suggested by Manaka *et al.* [5]

Figure 2 shows the field-swept spectra of the two samples $x$=0.85 and 1 at various temperatures under the field around 3T. At high temperatures both the samples exhibit a sharp paramagnetic resonance line. As decreasing temperatures, the $x$=1 sample keeps up the sharp width, while the $x$=0.85 sample shows a drastic split in the peak. At the lowest temperature of 4K, five distinct peaks appear overlapping each other, and $\Delta H$ defined as the full width at the 20% maximum of the entire spectrum as shown in Fig. 2, reaches around 1.2 kOe. The temperature dependence of $\Delta H$ is plotted in Fig. 3, where the finite value of $\Delta H$ above $T_N$ corresponds to the width of the paramagnetic peak $\Delta H_0$= 0.20 kOe.

This drastic splitting in the peak is the microscopic demonstration of the existence of the staggered field, and hence the antiferromagnetic long range order. The volume fraction of the ordered phase is considered to be around unity, because there is little spectral weight at the position at the paramagnetic peak. The temperature dependence of $\Delta H(T)-\Delta H_0$ in the ordered state is described as $(1-T/T_N)^{\beta}$ in the vicinity of the onset temperature of the peak splitting as shown in Fig. 3. The parameters of Néel temperature $T_N$ and the critical exponent $\beta$ were obtained from fitting to be 13.5(1) K and 0.33(1). The latter is consistent with 3D-Ising model, and coincides with the result of our μSR experiment [8].

Spectra taken under the various directions of the low field 0.8T are shown in Fig. 4 (a), where the applied field is rotated within A-plane. Positions of multiple peaks in observed spectra are qualitatively explained in terms of the inequivalent proton sites exposed to the hyperfine field produced by the staggered moment. Those ten inequivalent proton sites in the unit cell are expected to feel different hyperfine fields from electron spins, and hence to have different shifts. We have assumed the classical dipole-dipole interaction between electron and nuclear spins, and the magnetic structure of the simplest case $\vec{q}=0$, where every dimer orders antiferromagnetically with an identical phase to calculate $\delta H(\vec{r}_i)$, the hyperfine field at each proton site. $\delta H(\vec{r}_i)$ is evaluated from the simple formula $\delta H(\vec{r}_i) \approx \hat{h}_0 \cdot \sum_{j=1-N} A_{\rm dip}(\vec{r}_i-\vec{R}_j)\vec{\mu}(\vec{R}_j)$, where $\hat{h}_0$ is the unit vector along the applied field,

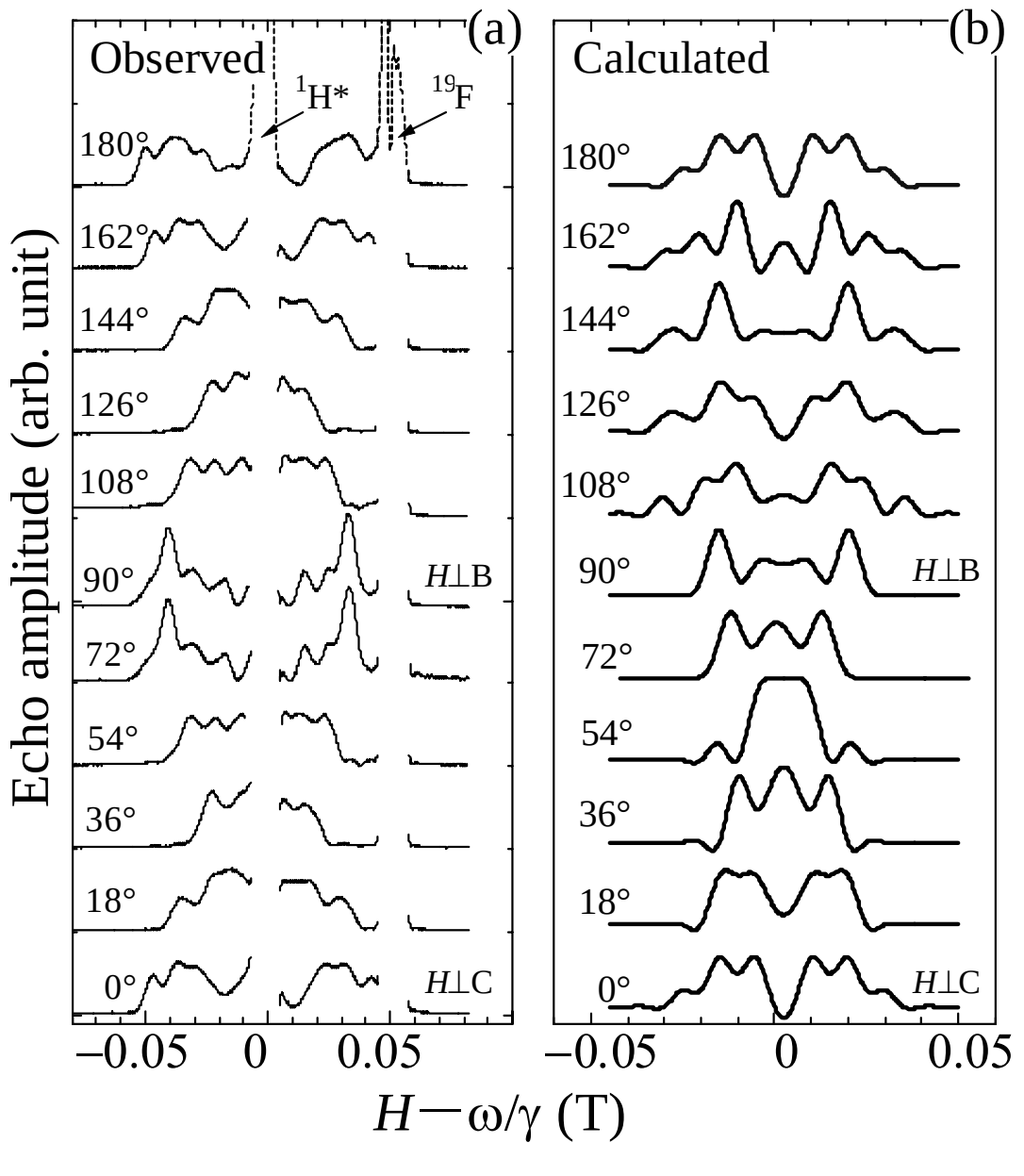

Fig. 4 (a) $^1$H-NMR spectra of the $x$=0.85 sample under 0.8T applied along various directions within A-plane. Some spurious signals $^1$H* and $^{19}$F are observed around the zero-shift position and +0.05T, indicated by arrows. (b) Calculated spectra based on assumptions of the classical dipole-dipole interaction, the ordered moment of 1$\mu_B$ directed perpendicular to C-plane, and the antiferromagnetic vector $\vec{q}=0$.

$A_{dip}$, the dipole tensor, $\vec{r}_i$ ($i$ =1-10), the proton sites in the unit cell, and $\vec{\mu}(\vec{R}_j)$, the magnetic moment located at the Cu site $\vec{R}_j$. $\vec{\mu}(\vec{R}_j)$ is assumed to be 1 $\mu_B$ in magnitude and lying along the easy axis perpendicular to C-plane. The latter is because the applied field 0.8T is much lower than the spin flip field around 2-3 T. In the calculation, summation on $j$ was taken over a hundred of neighboring Cu-sites. Obtained $\delta H(\vec{r}_i)$'s were convoluted with an appropriate Gaussian-width and superposed to obtain spectra shown in Fig. 4 (b). Though there are some differences in the detailed shape and the magnitude of the hyperfine field, one must note that the two show a good agreement in the overview, especially in the angle region 0-90°. This indicates the spin structure in this system is simple antiferromagnetic rather than an incommensurate or glass-like state, supporting Nakamura's theory [7].

For the complete determination of the antiferromagnetic vector, a theoretical investigation including the supertransferred hyperfine interaction via negative ions seems to be necessary. In fact, observed increase 1 kOe in $\Delta H$ from the paramagnetic value at lowest temperatures exceeds the calculated value 0.8 kOe expected for the ordered moment of 1$\mu_B$ with a dipole-dipole interaction, indicating the presence of the stronger interaction between electron spins and nuclear spins.

Finally, the knurled structure of spectra observed around $T_N$, shown in Fig. 2, does not mean an incommensurate spin state. Rather, it arose from an unavoidable small temperature instability of the system. This is confirmed simply by the fact that the two measurements under an identical condition did not coincide as shown by the two spectra #1 and #2 at 13.3K in Fig. 2. For the rising rate $\Delta H$ is very large around $T_N$, that is, $\Delta H/\Delta T \approx 0.1$T/K as shown in Fig. 3, a temperature drift within a few milli-Kelvin is enough to cause the observed fluctuation in spectra.

In conclusion, we have investigated $^1$H-NMR spectra on the bond-disordered spin gap system IPA-Cu($Cl_xBr_{1-x}$)$_3$ ($x$=0.85) to show that the ground state is antiferromagnetic rather than the glass-like or the gapped one. In order to investigate a change in the magnetism toward the quantum critical point, an experiment on samples with other concentrations is in progress.

**Acknowledgements**

The authors are grateful for kind support in specific heat measurements by PPMS to Dr. K. Noda at Kuwahara Lab., Sophia University, and magnetization measurements by MPMS to Prof. T. Nojima at Center of Low Temperature Science, Tohoku Univ. This work was supported by Kurata foundation and by Grant-in-Aid for Scientific Research on priority Areas "High Field Spin Science in 100T" (No.451) from the Ministry of Education, Culture, Sports, Science and Technology (MEXT).

**References**

1. H. Manaka, I. Yamada, and K. Yamaguchi, J. Phys. Soc. Jpn. **66**, 564 (1997).
2. H. Manaka and I. Yamada, J. Phys. Soc. Jpn. **66**, 1908 (1997).
3. T. Masuda, A. Zheludev, H. Manaka, L.-P. Regnault, J.-H. Chung and Y. Qiu, Phys. Rev. Lett. **96**, 047210 (2006).
4. F. D. Haldane, Phys. Rev. Lett. **50**, 1153 (1983).
5. H. Manaka, I. Yamada, H. Mitamura and T. Goto, Phys. Rev. **B66**, 064402 (2002).
6. H. Manaka, I. Yamada and H. Aruga Katori, Phys. Rev. **B63**, 104408 (2001)
7. T. Nakamura, Phys. Rev. **B71**, 144401 (2005).
8. T. Saito, A. Oosawa, T. Goto, T. Suzuki and I. Watanabe, Phys. Rev. **B**., *in press.* (cond-mat/ 0604466)